\documentclass[12pt]{article}
\usepackage{epsf}
\def\be{\begin{equation}}
\def\ee{\end{equation}}
\def\bea{\begin{eqnarray}}
\def\eea{\end{eqnarray}}

\begin{document}
\begin{titlepage}
\begin{center}
{\Large \bf William I. Fine Theoretical Physics Institute \\
University of Minnesota \\}
\end{center}
\vspace{0.2in}
\begin{flushright}
FTPI-MINN-13/22 \\
UMN-TH-3212/13 \\
July 2013 \\
\end{flushright}
\vspace{0.3in}
\begin{center}
{\Large \bf Suppression of the $S$-wave production of $\bigl ( {3 \over 2} \bigr ) ^+ + \bigl ( {1 \over 2} \bigr ) ^-$ heavy meson pairs in $e^+e^-$ annihilation
\\}
\vspace{0.2in}
{\bf  Xin Li$^a$  and M.~B.~Voloshin$^{a,b,c}$  \\ }
$^a$School of Physics and Astronomy, University of Minnesota, Minneapolis, MN 55455, USA \\
$^b$William I. Fine Theoretical Physics Institute, University of
Minnesota,\\ Minneapolis, MN 55455, USA \\
$^c$Institute of Theoretical and Experimental Physics, Moscow, 117218, Russia
\\[0.2in]

\end{center}

\vspace{0.2in}

\begin{abstract}
The heavy meson-antimeson pairs, where one is an excited $\bigl ( {3 \over 2} \bigr ) ^+ $ meson and the other is a ground state $\bigl ( {1 \over 2} \bigr ) ^-$ meson, namely the pairs ($D_1(2420) \bar D$ + c.c.), ($D_1(2420) \bar D^*$ + c.c.), ($D_2(2460) \bar D^*$ + c.c.) in the charm sector and ($B_1(5721) \bar B$ + c.c.), ($B_1(5721) \bar B^*$ + c.c.), ($B_2(5747) \bar B^*$ + c.c.) in the bottom sector, are allowed, by the quantum numbers, to be produced in the $S$ wave in $e^+e^-$ annihilation. We show, however, that such $S$-wave production is forbidden by the heavy quark spin symmetry. Thus the yield of the considered meson pairs in $e^+e^-$ annihilation should be significantly suppressed near the respective thresholds. In our view, this substantially weakens the arguments for considering the $Y(4260)$ charmonium-like resonance as a $D_1 \bar D$ molecular state. 
\end{abstract}
\end{titlepage}

In the orbitally excited heavy $Q \bar q$ mesons ($D_1(2420)$ and $D_2(2460)$ with $Q=c$ and $B_1(5721)$ and $B_2(5747)$ with $Q=b$) the light (anti)quark is in the state with the quantum numbers $J^P=(3/2)^+$. The heavy quark spin symmetry (HQSS) thus requires that in their decay into the ground state heavy mesons and a pion, e.g. $D_1(2420) \to D^* \pi$, $D_2(2460) \to D \pi$, $D_2(2460) \to D^* \pi$, the pion is emitted in the $D$-wave. For this reason these excited heavy mesons are relatively narrow, having the width of about 25\,MeV, unlike the other pair of the orbitally excited mesons, where the light antiquark is in the $J^P=(1/2)^+$ state, e.g. $D_0(2400)$ and $D_1(2430)$, whose widths are in the ballpark of 300\,MeV due to similar decays with the pion emission in the $S$-wave~\cite{pdg}. The $J^P=(3/2)^+$ mesons can be combined with the ground-state (anti)mesons in the $S$-wave to form pairs with the quantum numbers $J^{PC}=1^{--}$ matching those necessary for direct production in the $e^+e^-$ annihilation. This property has lead to the suggestion~\cite{ding,lwdz,whz,ghmwz,liuli} that the resonance $Y(4260)$ observed in the $e^+e^-$ annihilation is a near-threshold $S$-wave bound `molecular' state of ($D_1(2420) \bar D$ - c.c.) charmed meson pair. If correct, this picture would generally imply, due to the HQSS, an  existence of a whole family of similar threshold charmonium states with either the $D$ meson replaced by $D^*$ or the $D_1(2420)$ replaced by $D_2(2460)$. Some of these charmonium-like states with $J^{PC}=1^{--}$ should be directly observable as resonances in the $e^+e^-$ annihilation at the total energy (4.3 - 4.4)\,GeV, namely molecular states of $D_2(2460) \bar D^*$ and $D_1(2420) \bar D^*$. Moreover a similar `suite' of bottomonium-like molecular resonances would be expected in $e^+e^-$ annihilation near 11.0\,GeV.
  
The purpose of this paper is to show that in fact the production of the discussed $J^{PC}=1^{--}$ $S$-wave pairs with one orbitally excited $(3/2)^+$ meson is forbidden by the HQSS. This conclusion casts doubt on the interpretation of the resonance $Y(4260)$ as a $D_1 D$ molecular state, since it effectively removes the important argument that being an $S$-wave state it carries no significant threshold suppression for its yield in the $e^+e^-$ annihilation. Although, indeed there is no kinematical suppression, the production amplitude is suppressed by the inverse of the heavy quark mass, which is the parameter for breaking the HQSS. For the charmed quark one would generally expect the latter suppression to amount to a factor of order 0.1, or stronger, in the rate, and a significantly stronger suppression for the heavier bottom quark sector. In view of our result, an alternative interpretation of the structure of $Y(4260)$ resonance, e.g. as hadro-charmonium~\cite{mv07,dv}, may be more credible.

The claimed in this paper selection rule for production of the heavy meson pairs is strictly valid in the limit of HQSS. Indeed, in this limit the spin of both the heavy quark and the antiquark is strictly conserved. In other words, in the conversion of the heavy quark-antiquark pair $\bar Q \, Q$ produced by the electromagnetic current into the final state of heavy mesons the spin of the heavy quarks is not dynamical, and one can consider instead spinless heavy quarks. Then the considered orbitally excited mesons have the quantum numbers $J^P=(3/2)^+$, and the ground-state heavy mesons have $J^P=(1/2)^-$. The electromagnetic current $(\bar Q \gamma_\mu Q)$ for slow heavy quarks has the nonrelativistic form $\bar Q \sigma_i Q$ and couples only to the spin of the heavy quarks. Once the spin of the heavy quarks is removed, the equivalent operator generating the heavy quark pair is equivalent to a unit operator (with the point-like spatial structure $\delta^3(\vec r)$) corresponding to the quantum numbers $J^{PC}=0^{++}$. Clearly, it is impossible to make the total spin 0 state out of an $S$-wave pair of mesons with spin 3/2 and 1/2. Thus no $S$-wave state can be produced, and this conclusion is valid for any combination of the polarizations of the mesons. In the real situation, where the heavy quarks have spin, these polarization states combine with the spin of the $\bar Q Q$ pair to form the discussed here three types of the meson-antimeson pairs with $J^{PC}=1^{--}$. Thus no such pair can be produced in the $S$ wave. 

An alternative proof of our conclusion, where the spin of the heavy quarks is not removed and is explicitly traced, goes as follows. A state of the heavy meson pair can be decomposed~\cite{bgmmv,mv11,mv12} in terms of the total spin of the $\bar Q \, Q$ pair, $S_H$, and the total angular momentum $S_L$ of the rest (`light') degrees of freedom, where the latter includes both the total spin of the light quark pair $\bar q \, q$ and any orbital momentum. The states with $J^{PC}=1^{--}$ are generally a combination of four eigenstates of the operators $\vec S_H$ and $\vec S_L$:  
\be
\psi_{10}=1^{--}_H \otimes 0^{++}_{L}\,, ~~~\psi_{11}=1^{--}_H \otimes 1^{++}_{L}\,,~~~ \psi_{12}=1^{--}_H \otimes 2^{++}_{L}\,, ~~{\rm and}~~ \psi_{01}=0^{-+}_H \otimes 1^{+-}_{L}\,.
\label{psis}
\ee
Clearly, in the $(3/2)^+$ mesons the light degrees of freedom carry angular moment 3/2, while in the ground state mesons they carry the angular momentum 1/2. Thus combining two such mesons in the $S$-wave would never produce a state with $S_L=0$, and the component $1^{--}_H \otimes 0^{++}_{L}$ should be absent in the decomposition of the meson pair states in terms of $S_H \otimes S_L$ eigenstates. This can be also readily verified by performing the transformation explicitly:
\bea
(D_1 \bar D - \bar D_1 D)~ &:&~~~ {1 \over 2 \sqrt{2}} \, \psi_{11} + {\sqrt{5} \over 2 \sqrt{2}} \, \psi_{12} + {1 \over 2} \, \psi_{01} \nonumber \\
(D_1 \bar D^* - \bar D_1 D^*) &:&~~~ {3 \over 4} \, \psi_{11} - {\sqrt{5} \over 4} \, \psi_{12} + {1 \over 2 \sqrt{2}} \, \psi_{01} \nonumber \\
(D_2 \bar D^* - \bar D_2 D^*) &:&~~~ {\sqrt{5} \over 4} \, \psi_{11} + {1 \over 4} \, \psi_{12} - {\sqrt{5} \over 2 \sqrt{2}} \, \psi_{01}
\label{trans}
\eea
(It can be noticed that the meson-pair wave functions and the functions $\psi_{ij}$ are assumed to be normalized to one, so that the transformation matrix is orthogonal.) The missing $S_H \otimes S_L$ eigenstate $\psi_{10}$ is exactly the one (and the only one) produced by the electromagnetic current $(\bar Q \sigma_i Q)$. Thus the production of each of the states in Eq.(\ref{trans}) is forbidden in the limit of HQSS where $S_H$ is conserved.

One might argue that the HQSS breaking  can be enhanced in the threshold region~\cite{mv11}, where the mass splitting between the mesons related by this symmetry is comparable to the distance to the thresholds. We can note however, that this is very unlikely to be the case for our selection rule. Indeed at the threshold for each of the considered meson-antimeson channel the major effect is the mass splitting $\mu$ between the $D^*$ and $D$ mesons, $\mu \approx 140\,$MeV, which is approximately the same as the energy gap between the thresholds. This could generally lead to a significant violation of HQSS due to rescattering between the channels. However in the discussed case {\it neither} of the three channels can be produced, so that the rescattering between them cannot give rise to a non-zero production amplitude. The distance to the other thresholds is parametrically of order $\Lambda_{QCD}$, so that the effects of the HQSS breaking can be estimated as being of order $\mu/\Lambda_{QCD}$.

The latter behavior can be illustrated by considering a simple model for rescattering of the type $\gamma^* \to D^* \bar D^* \to D_1 \bar D$ due to the pion exchange as shown in Fig.~1. (One can readily verify that the intermediate channels with at least one pseudoscalar $D$ meson instead of the vector $D^*$ meson cannot scatter into $D_1 \bar D$.) For the $S$-wave final state $D_1 \bar D$ one can set the momentum of the final mesons to zero, so that in the diagram of Fig.~1 there is only the loop momentum $\vec p$. The amplitude for the production of the $D^* \bar D^*$ pair in $e^+e^-$ annihilation can be written in the general form~\cite{lv} in terms of three partial wave amplitudes: 
\bea
&&A(e^+e^- \to D^* \bar D^*) = A_0(p^2) \, j_k p_k \cdot {1 \over 3} \, a_l b_l + j_k \cdot {1 \over \sqrt{20}} \left ( a_i b_j +a_j b_i - {2 \over 3} \delta_{ij} \, a_l b_l \right ) \times \nonumber \\
&& \left \{ A_2(p^2) \,  \delta_{ki} p_j+ {5 \over \sqrt{6}} \, A_F(p^2) \, \left [ {1 \over p^2} \, p_i p_j p_k - {1 \over 5} \, \left (p_k \delta_{ij}+p_j \delta_{ik} + p_i \delta_{kj} \right ) \right ] \right \}~,
\label{3amp}
\eea
where $\vec a$ and $\vec b$ are the polarization amplitudes for the meson and anti-meson and $\vec j$ denotes the polarization amplitude of the virtual photon. The amplitudes $A_0$ and $A_2$ are the $P$-wave amplitudes corresponding to the production of the meson pair with the respective total spin $S=0$ and $S=2$, and $A_F$ is standing for the $F$-wave one. The relative normalization of the amplitudes in Eq.(\ref{3amp}) corresponds to the production cross section being proportional to $p^3 \, (|A_0|^2 + |A_2|^2 + |A_F|^2)$. One can also notice that under this normalization the expansion of $A_F$ at small momentum $p$ starts with $p^2$.

\begin{figure}[ht]
\begin{center}
 \leavevmode
    \epsfxsize=10cm
    \epsfbox{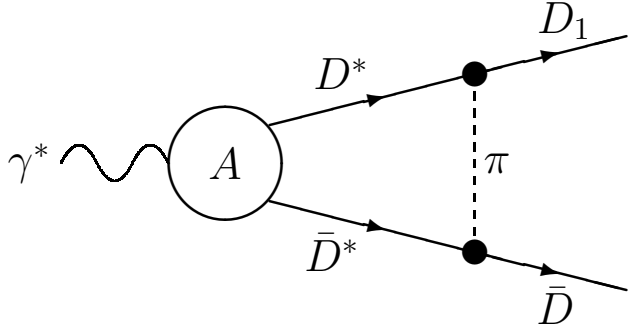}
    \caption{Rescattering through pion exchange of $D^* \bar D^*$ meson pair produced in $e^+e^-$ annihilation into $D_1 \bar D$}
\end{center}
\end{figure}

In the HQSS limit  one has~\cite{kmm,mv11,lv}
\be
A_F=0,~~~~{\rm and}~~~A_2=- 2 \sqrt{5}\,A_0~,
\label{hqss}
\ee
so that the tensor structure of the production amplitude is in fact given by
\be
A(e^+e^- \to D^* \bar D^*)= - A_0 \, j_i p_j \,  \left ( a_i b_j +a_j b_i -  \delta_{ij} \, a_l b_l \right )~.
\label{hqssa}
\ee
Using this structure and the required $D$-wave coupling $D_1 D^* \pi$: $\propto (D_1)_i (D^*)_j \, (3 \, p_i p_j - \delta_{ij} \, p^2)$, and the $P$-wave coupling $D^* D \pi$: $\propto (D^*)_i p_i$, one readily finds that the tensor structure of the loop integral for the diagram of Fig.~1 is proportional to $j_i (D_1)_j \, (3 \, p_i p_j - \delta_{ij} \,p^2) \, p^2$ and vanishes upon the integration over the direction of the loop momentum $\vec p$. Clearly, this cancellation is due to the symmetry relations (\ref{hqss}).

The relations (\ref{hqss}) generally are not expected to hold near the $D^* \bar D^*$ threshold~\cite{mv11} due to the enhanced effect of the mass splitting $\mu$. In particular, they are broken there by the enhanced partial wave mixing arising from the rescattering within the meson pairs~\cite{lv}. However well above the $D^* \bar D^*$ threshold the effects of the mass splitting become small and the HQSS is restored. In particular, as previously mentioned, these effects should be parametrically of order $\mu/\Lambda_{QCD}$ at the $D_1 \bar D$ threshold. One can also readily verify a similar cancellation for the $S$-wave production of $D_1 \bar D^*$ and $D_2 \bar D^*$ by rescattering through the intermediate channels $D^* \bar D^*$, $D^* \bar D$ and $D \bar D$, provided that the production amplitudes for the latter channels satisfy the HQSS relations.

Another effect of HQSS breaking could be a mixing between the orbitally excited $(3/2)^+$ and $(1/2)^+$ spin 1 mesons, e.g. a mixed structure of the $D_1(2420)$ and $D_1(2430)$. However, phenomenologically this mixing is apparently very small, as one can deduce from the small total width of $D_1(2420)$ as compared to that of $D_1(2430)$. Had the mixing been substantial, it would have introduced a large $S$-wave component in the amplitude of the decay $D_1(2420) \to D^* \pi$ resulting in a considerable enhancement of the rate of this decay.

We thus conclude that the yield in the $e^+e^-$ annihilation of the discussed $(3/2)^+ + (1/2)^-$ $S$-wave pairs of heavy mesons, forbidden in the limit of HQSS, should be significantly suppressed when the effects of the spin symmetry breaking are also taken into account.

MBV thanks A.~Bondar for numerous illuminating discussions.
This work is supported, in part, by the DOE grant DE-FG02-94ER40823.

\end{document}